# Ultralight and ultra-stiff nano-cardboard panels: mechanical analysis, characterization, and design principles

Jong-hyoung Kim[1], Lishuai Jin[1], Benjamin C. Schafer[1], Quan Jiao[1], Katia Bertoldi[1], David W. Keith[1], Joost J. Vlassak[1,*]

**Affiliations**

[1]*John A. Paulson School of Engineering and Applied Sciences, Harvard University, MA, 02138, USA*

[*]Address all correspondence to this author. E-mail: vlassak@seas.harvard.edu

**Keywords**

Nanoarchitected material, nano-cardboard, sandwich panel, mechanical behavior, buckling

**Abstract**

    We introduce a class of ultra-light and ultra-stiff sandwich panels designed for use in photophoretic levitation applications and investigate their mechanical behavior using both computational analyses and micro-mechanical testing. The sandwich panels consist of two face sheets connected with a core that consists of hollow cylindrical ligaments arranged in a honeycomb-based hexagonal pattern. Computational modeling shows that the panels have superior bending stiffness and buckling resistance compared to similar panels with a basketweave core, and that their behavior is well described by Uflyand-Mindlin plate theory. By optimizing the ratio of the face sheet thickness to the ligament wall thickness, panels maybe obtained that have a bending stiffness that is more than five orders of magnitude larger than that of a solid plate with the same area density. Using a scalable microfabrication process, we demonstrate that panels as large as 3x3 cm$^2$ with a density of 20 kg/m$^3$ can be made in a few hours. Micro-mechanical testing of the

panels is performed by deflecting microfabricated cantilevered panels using a nanoindenter. The experimentally measured bending stiffness of the cantilevered panels is in very good agreement with the computational results, demonstrating exquisite control over the dimensions, form, and properties of the microfabricated panels.

1. Introduction

Sandwich panels are structural elements that consist of a lightweight foam or honeycomb core between two thin face sheets. They are commonly used in aviation, transportation, and construction where mechanical performance and weight saving are critical. Here, we explore the use of sandwich panels at an entirely different length scale. The goal is to create small planar structures that are light enough to be lofted by photophoretic forces, but that have sufficient rigidity and strength to be handled at a macroscopic scale and to potentially serve as airborne substrates for small payloads in the Earth's stratosphere. Photophoretic forces arise as a result of molecular interactions when a structure exposed to light develops a non-uniform temperature distribution in a rarefied atmosphere[1]. While small, photophoretic forces caused by solar radiation may be sufficient to levitate very light structures in the Earth's stratosphere [1, 2]. The advantage of using photophoretic forces for this purpose is that such a structure can stay within a certain altitude range semi-permanently [3] where it may serve as a platform for atmospheric sensors and communication devices on Earth or even Mars [4].

Ultra-light sandwich panels at the length scales required for photophoretic lofting have only recently gained attention. Monolithic sandwich panels with cores that consist of pyramidal or

---

[1] Crookes' radiometer, which consists of a set of vanes on a low-friction spindle in a partially evacuated glass bulb, is a well-known illustration of the photophoretic force. When illuminated, the vanes rotate with an angular velocity that depends on the intensity of the light.

octahedral truss structures have been fabricated by first making a sacrificial polymer template using two-photon lithography, coating this template with electroless Ni-P, and then finally etching away the sacrificial template in an alkaline solution [5, 6]. Typical sandwich panels made using this technique have a panel thickness of 1 mm and a Ni-P coating thickness in the range of 1 μm. Mechanical characterization of these panels shows that they are quite robust, but the panels have a fairly large area density in the range of 100-1000 g/m$^2$, mainly due to the relatively large coating thickness and the high density of Ni-P. This density is too large for photophoretic lofting under atmospheric insolation [1, 2]. In order to fabricate sandwich panels with much lower area densities, Bargatin *et al.* proposed a cardboard-like sandwich panel with a basketweave core fabricated using conventional silicon microfabrication techniques [7, 8]. These panels are made by first patterning the silicon layer of a silicon-on-insulator (SOI) wafer using photolithography and deep reactive-ion etching (DRIE) to form a sacrificial template. This template is then coated with a thin layer of alumina using atomic layer deposition (ALD). In a final step, the sacrificial silicon is etched away using an isotropic etch, leaving a sandwich panel with a thickness in the range of 10 μm and an alumina thickness on the order of 50 nm. These panels have an area density of only 0.5 g/m$^2$ and a volumetric density of 50 kg/m$^3$, on par with a typical silica aerogel [9, 10], yet are sufficiently robust to be handled with tweezers. Unfortunately, use of SOI wafers as sacrificial templates limits the maximum panel thickness that can be achieved, and the fabrication process is not readily scalable to panels larger than approximately 1 cm$^2$. Because of their superficial similarity to cardboard, we follow Bargatin *et al.* in referring to microfabricated sandwich panels as nano-cardboard [1, 8]

Here we investigate the mechanical behavior of a range of microfabricated sandwich panels, both using computational analyses and micro-mechanical testing. We propose a novel hexagonal unit-cell geometry that results in isotropic in-plane properties and that has superior buckling resistance. We evaluate the bending stiffness, shear stiffness, and post-buckling behavior for a broad range of unit cell geometries and dimensions and use these results to formulate guidelines for the design of useful sandwich structures. We describe a scalable microfabrication process to fabricate larger sandwich panels much more quickly than is possible with existing process flows. Finally, we perform micro-mechanical tests on microfabricated cantilevered panels and find very good agreement between the experimental bending stiffness and the computational results, demonstrating exquisite control over the dimensions, form, and properties of the microfabricated panels.

## 2. Fabrication and Characterization

### 2.1. Geometry of the unit cell

Figure 1a shows a schematic drawing of a nano-cardboard panel with a core that consists of cylindrical elements organized in a periodic pattern. The geometry of the unit cell used for the panel is an important factor in determining both the area density of the panel and its mechanical behavior. Because the face sheets of nano-cardboard panels are so thin, the bending stiffness of the individual face sheets is exceedingly small. This small bending stiffness may lead to spontaneous wrinkling of the face sheets during the fabrication process. Lin and Bargatin [8] proposed a no-straight-line condition to prevent spontaneous wrinkling in the face sheet.

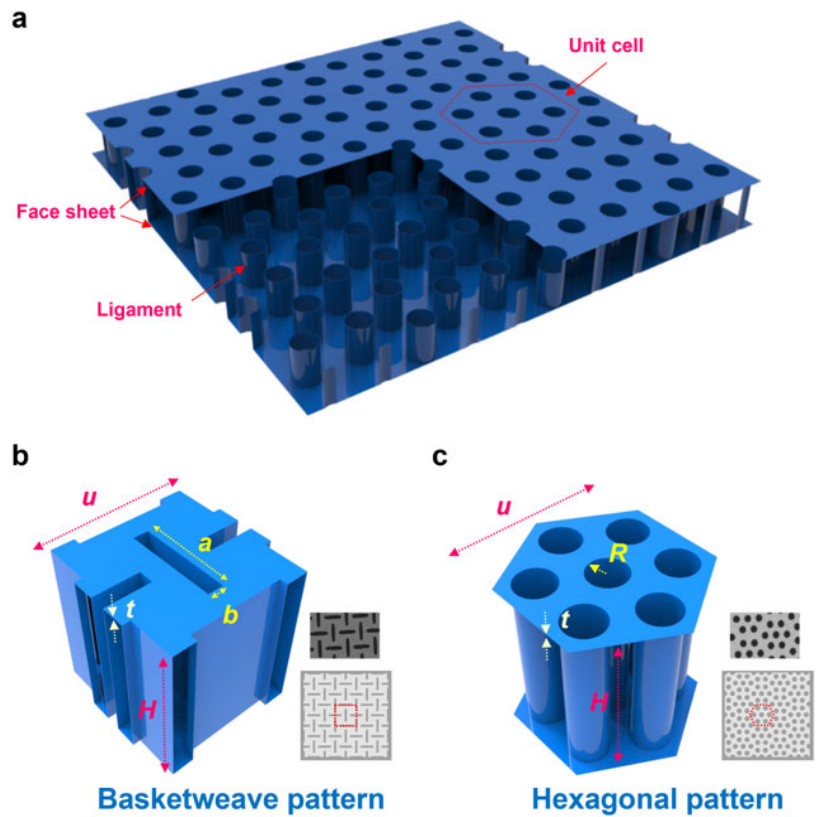

**Figure 1.** (a) Schematic diagram of 2D nanolattice structures with hexagonal pattern; (b) Unit cell of basketweave pattern; (c) Unit cell of hexagonal pattern.

According to this condition, the unit cell of a sandwich panel needs to be defined such that it is impossible to draw a straight line across the face of the panel without intersecting a ligament connecting the two face sheets. The basketweave pattern depicted in Figure 1b satisfies this condition and was studied extensively by these researchers. While the basketweave pattern has a large specific bending stiffness, its elastic properties are not isotropic in the plane of the panel because the unit cell has only four-fold rotational symmetry [8].

Here, we introduce a honeycomb-like unit cell that contains a pattern of hollow cylindrical ligaments with hexagonal rotational symmetry (Figure 1c). Like the basketweave pattern, this pattern satisfies the no-straight-line condition, but it results in a panel that has in-plane elastic isotropy because of its rotational symmetry (*Appendix A*). Furthermore, for the same wall thickness and cross-sectional area, hollow cylinders with circular cross-section have critical buckling loads that are orders of magnitude larger than cylinders with rectangular cross-section, leading to better overall performance of the panel.

## 2.2. Nano-cardboard fabrication process

Microfabrication of a sandwich panel consists of three distinct steps 1) the formation of a sacrificial template, 2) deposition of a thin conformal layer onto the sacrificial template, and 3) selective removal of the sacrificial template to leave a freestanding structure [8, 11]. We first fabricate a sacrificial template from a general-purpose, single-crystal silicon substrate (Figure 2a). The thickness of the substrate determines the thickness of the sandwich panel, while the size of the substrate sets an upper limit on the in-plane dimensions of the panel. We use standard photolithography to define the ligament pattern on the silicon substrate and then use a deep reactive ion etch (DRIE) based on the Bosch process to etch all the way through the silicon substrate. Once the ligament pattern has been etched, we coat the silicon template with a thin layer of aluminum oxide. To obtain a coating of uniform thickness independent of the geometry of the template and across the entire silicon substrate, we use the atomic layer deposition process (ALD), which deposits the coating atomic layer by atomic layer over the entire extent of the template. Coatings with thicknesses ranging from tens to hundreds of nanometers are readily achieved using this process. We later refer to the thickness of the ALD coating as the wall thickness since the ALD coating forms both the face sheets and the ligament walls of the sandwich panel. For the few cases where the thicknesses of the face sheets and the ligaments are different, we will refer to these values explicitly as face sheet thickness ($t_f$) or ligament wall thickness ($t_l$). Different thicknesses of the face sheets or ligament walls can be achieved with additional deposition or etch processes.

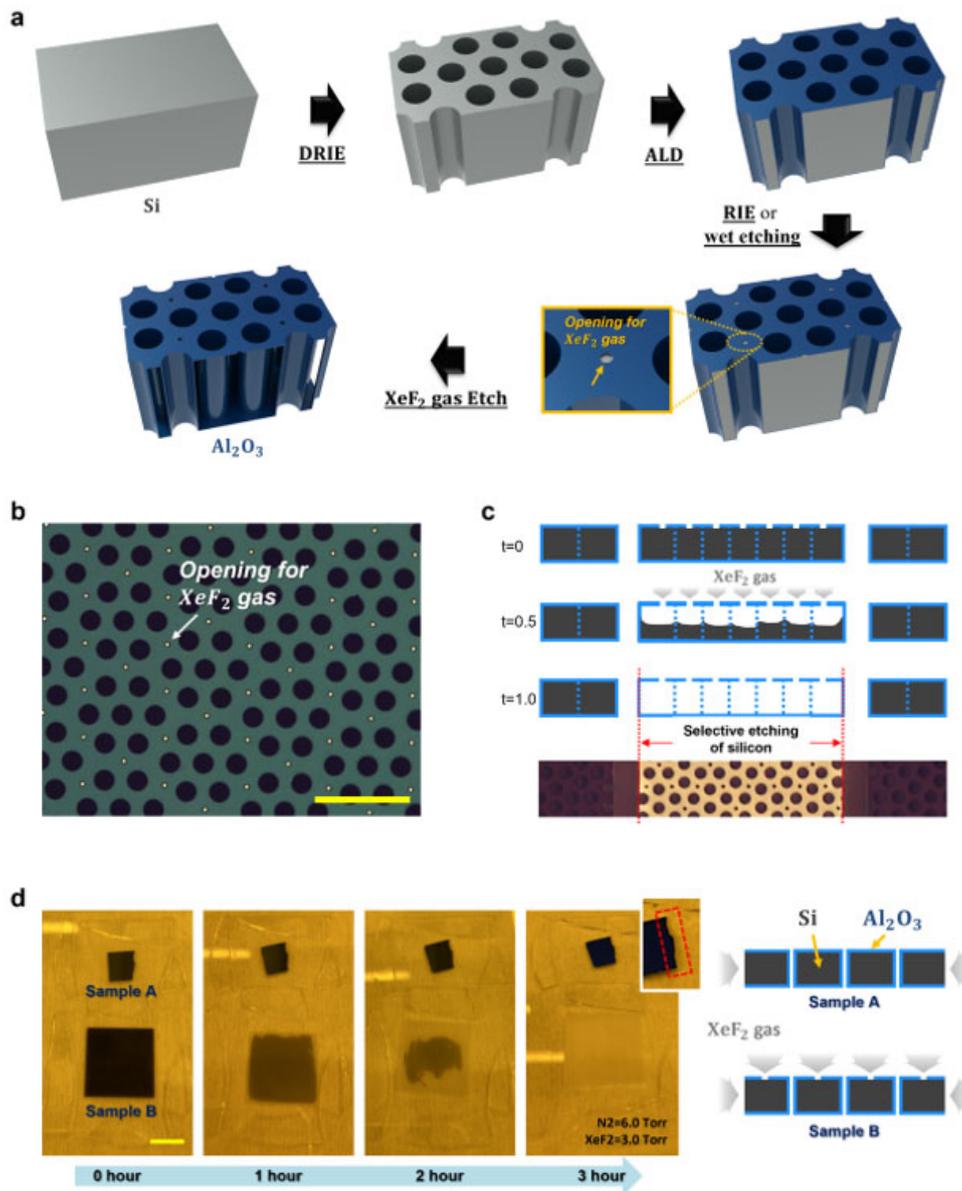

**Figure 2.** (a) Schematic diagram of the process used to fabricate nano-cardboard panels; (b) Optical image of a sample with micron-size openings in the top layer immediately prior to the $XeF_2$ etching process; (c) Selective etching of silicon using $XeF_2$ gas by spatially distributing the openings in the top layer; (d) Effect of the openings in the top layer on how quickly the sacrificial template is etched for the case of a 100 μm thick silicon template. The scale bar corresponds to 100 μm in figure (b) and to 10 mm in figure (d).

In a final step, an isotropic $XeF_2$ etch selectively removes the sacrificial silicon template while leaving the aluminum oxide in place. Because the silicon is completely encapsulated in aluminum oxide, it is necessary to first introduce small openings in the aluminum oxide to provide access to the silicon (Figure 2b). These openings are defined by means of photolithography using a photoresist (SPR 220-7.0) that bridges the openings in the silicon template. The aluminum oxide is then etched using either a buffered oxide (5:1) etch solution, or a reactive ion etch based on a $BCl_3/Cl_2/Ar$ chemistry. Since the size of these openings is very small compared to the size of the unit cell, they can be uniformly distributed across the template without reducing the mechanical performance of the panels in a meaningful way (*Appendix B*). The openings in the aluminum oxide allow the $XeF_2$ gas to etch the silicon uniformly across the silicon template (Figure 2c), radically decreasing the time necessary to remove the silicon and making the etch time independent of the size of the template. Figure 2d illustrates the effectiveness of this approach. *Sample B* in Figure 2d has small openings in the top aluminum oxide layer to allow uniform access of $XeF_2$, while *Sample*

*A* only provides access via the edges of the template as in the process described by Lin and Bargatin [8]. It is clear that all the silicon in *Sample B* has been etched after a few hours, while only a small amount of silicon has been removed from *Sample A* in the same amount of time, even though *Sample B* is more than four times larger than *Sample A*. Additionally, this approach makes it possible to selectively etch away silicon in certain sections of the template, while leaving it in place in others, simply by controlling the spatial distribution of the openings in the top layer (Figure 2c). This feature of the fabrication process enables fabrication of nano-cardboard panels of a variety of shapes, including nano-cardboard panels tethered to a solid silicon frame, and also provides an opportunity to integrate these structures into MEMS devices. Figure 3 shows a few examples of nano-cardboard panels resulting from this fabrication process. Macro-scale panels are readily fabricated (Figure 3a) and are rigid enough to be handled by hand or tweezer (Figure 3b).

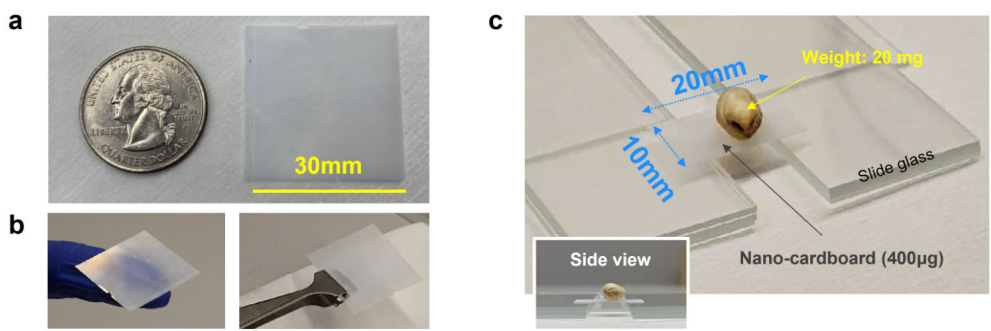

**Figure 3.** (a) Macro-scale nano-cardboard panel larger than a US quarter coin; (b) Handling of macro-scale nano-cardboard panel by hand and tweezer; (c) A nano-cardboard panel fabricated using a 100 nm ALD coating supports a mass of 20 mg. The area density of the nano-cardboard in this figure is 2.0 g/m²

The panel in Figure 3c supports a mass of 20 mg without noticeable deflection or fracture.

### 2.3. *Computational analysis*

The mechanical behavior of the nano-cardboard panels was analyzed using both computational and experimental techniques. The computational analysis was performed using a commercial finite-element analysis software package, ABAQUS 2019, for several different unit cells and for various values of the ALD coating and panel thicknesses. To examine the bending behavior of the panels, we applied either a moment (pure bending) or a transverse force (combined bending and shear) at the end of an infinitely wide clamped cantilever plate made out of nano-cardboard (Figure 4a). To simulate the infinitely wide plate, periodic boundary conditions were applied to a cantilever beam that was a single unit cell wide [12]. For ease of modeling and application of the boundary conditions, we represent the honeycomb structure using the rectangular unit cell shown in Figure 4a instead of the hexagonal unit cell in Figure 1c [13, 14]. We also examined the behavior of a single unit cell subject to simple shear (Figure 4b) [15]. In this case, the unit cell was fixed on one side and a transverse force was applied to the nodes on the opposite side with boundary conditions to prevent rotation of the section; periodic boundary conditions were applied to the nodes on both side faces. The nano-cardboard was modeled using the general-purpose shell element S4R, because the wall of the nano-cardboard consists of an ALD coating that is typically two to four orders of magnitude smaller than the overall macroscopic dimensions of the panel. The number of elements in a single unit cell varied from 7,800 to 32,500 depending on its geometry. We used a linear elastic model with Young's modulus of 170 GPa and a Poisson's ratio of 0.21, typical values for ALD coatings of aluminum oxide [16, 17]. The length of the cantilever beams ($L$) was as indicated in the various figures. The simulations were performed with large-displacement kinematics and included extensive post-buckling behavior.

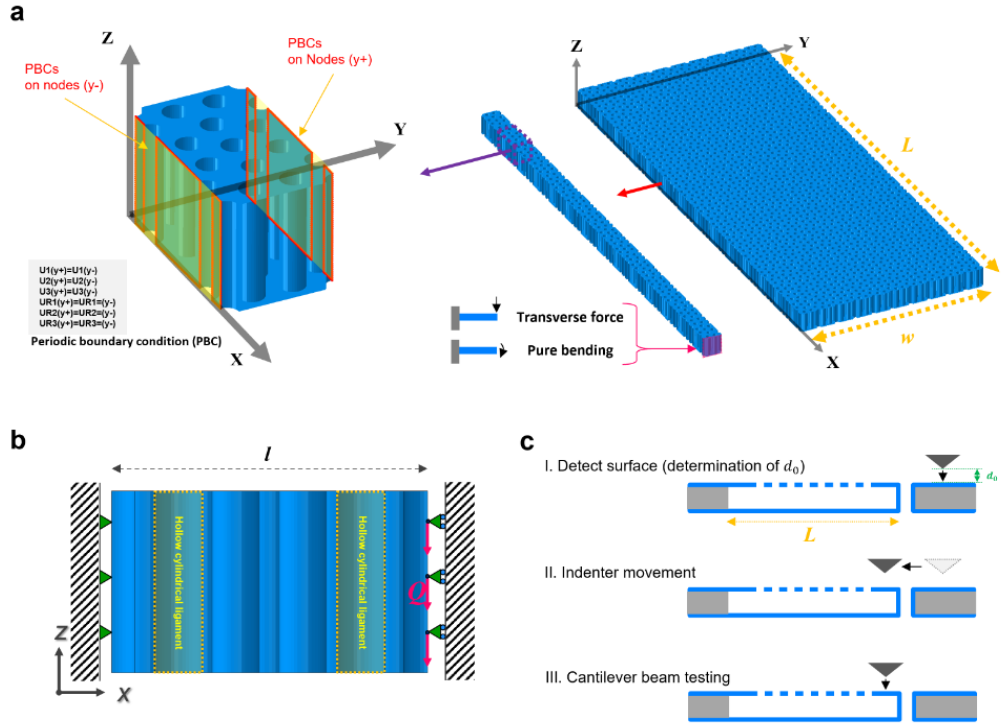

**Figure 4.** (a) Infinitely wide plate modeled as a cantilever beam with a width of one unit cell and periodic boundary conditions. Unit cell width cantilever beam for analyzing various force modes using finite element analysis and surface-coupling implementation of PBCs; (b) Boundary condition and applied load on the structure for simulating simple shear; (c) Schematic diagram of nanoindentation procedure for 2D nanolattice cantilever beam testing.

## 2.4. Experimental characterization

The mechanical behavior of the nano-cardboard panels was also measured experimentally. Nano-cardboard cantilever panels with a hexagonal unit cell and ligaments with a 13 μm radius ($R$) were first fabricated using the microfabrication process described earlier. The length of the cantilever panels was either 850 ± 0.4 μm or 1000 ± 0.4 μm, and their width was 810 ± 0.4 μm. The thickness of the panels was 105 ± 5 μm; the thickness of the aluminum oxide used to make the panels was determined to be 320 ± 5 nm by observing the cross-section of the cantilever panel in a scanning electron microscope (SEM). The load-deflection curves of the cantilever panels were measured using a nanoindentation tester (Nanomechanics, Inc., USA) with a Berkovich indenter tip, imposing a maximum displacement of 7 μm. To avoid damage to the nano-cardboard and to better determine the initial point of contact between the indenter and the cantilever beam, we used a customized testing sequence, shown in Figure 4c. First, the initial distance between the indenter and the cantilever panel was determined by making an indentation in the rigid area where the silicon had not been etched away, close to the tip of the cantilever beam. Then the indenter was positioned above the tip of the cantilever panel, a distance 50 μm from the edge, and the indenter was lowered slowly to measure the load-deflection curve of the cantilever.

## 3. Results and discussion

### *3.1. Initial bending stiffness and shear stiffness of nano-cardboard panels*

### *Dependence on wall thickness (t) and panel thickness (H)*

Like their macroscopic counterparts, nano-cardboard panels consist of two face sheets and a lightweight core. Nano-carboard panels, however, have three structural features that are quite different from macroscopic panels: 1) The volumetric fraction of material in the core is extremely low, typically on the order of $10^{-3}$, 2) the ligaments between the face sheets are not connected to each other in the plane of the panel other than by the face sheets, and 3) the ratio of the thickness of the face sheet (i.e., the ALD coating thickness) to the thickness of the panel (i.e., the silicon substrate thickness) is very small, typically $10^{-2}$ to $10^{-4}$. As illustrated below, these characteristics result in somewhat different bending and shear behavior compared to conventional panels. We have calculated the initial bending and shear stiffnesses of nano-cardboard panels as a function of wall thickness $t$ and panel thickness $H$, while keeping the unit cell of the panel the same (hexagonal pattern $R/u$=0.12, where $R$ and $u$ are defined in Figure 1c). The bending stiffness per unit width of the nano-cardboard panel, $K_B = ML/\varphi$, is calculated as the initial slope of the moment-rotation curve obtained under conditions of pure bending. Here, $M$ is the moment per unit width applied at the end of a cantilever plate of length $L$, and $\varphi$ is the angle of rotation of the end of cantilever plate. The results are summarized in Figure 5a. The bending stiffness of the nano-cardboard panels is proportional to $t \cdot H^2$, as one would expect for a sandwich panel with $t/H \rightarrow 0$ and with a core of negligible bending stiffness. In this limit, linear sandwich theory predicts a bending stiffness $K_B$ (*Appendix C*) [18, 19],

$$K_B \approx E_f \frac{t \cdot H^2}{2}, \qquad \text{Eq. (1)}$$

where $E_f$ is the homogenized elastic modulus of the face sheets. The value of $E_f$ for a nano-cardboard panel with a hexagonal pattern can be estimated directly by dividing the in-plane stiffness of the panel by a factor of two. The dotted curves in Figure 5a represent the bending stiffness calculated from Eq. (1) using this estimate and are in very good agreement with the finite element model (FEM) calculations.

The shear stiffness per unit width ($K_S$) of the nano-cardboard panel was determined from the deformation of a rectangular unit cell subjected to simple shear (Figure 4b), $K_S = Ql/\Delta x$, where $\Delta x$ is the transverse displacement, $Q$ the applied force per unit width, and $l$ the length of unit cell. The results are depicted in Figure 5b.

The shear stiffness of the nano-cardboard panels is independent of panel thickness and approximately proportional to $t^3$. At first glance, this result is somewhat surprising since the shear stiffness of a conventional macroscopic sandwich panel is typically well approximated by the shear stiffness of the core. In the case of these nano-cardboard panels, the shear stiffness of the core is relatively large because the core consists of cylindrical shells that serve as ligaments between the face sheets. The dependence of the shear stiffness on thickness of the face sheets then arises because the shear deformation imposed on the unit cell is accommodated almost entirely by bending of the face sheets in the regions around the cylindrical ligaments rather than by shear deformation of the core. The cylindrical ligaments are so stiff that they are not the limiting components for shear deformation, and because they are not directly connected to each other, the ligaments constrain the deformation of the face sheet less than if they were connected. This mode of deformation is illustrated in Figure 5c, which shows a cross-section through a small nano-cardboard panel deformed in simple shear. It is evident from the figure that the face sheets are corrugated, while the cylindrical ligaments are virtually undeformed. Since the deformation of the panel is accommodated by local bending of the face sheets, the overall shear stiffness is determined by the bending stiffness of the individual face sheets, i.e., $K_S \sim t^3$. The dependence of $K_S$ on $t$ is shown in more detail in Figure 5d over several orders of magnitude of the wall thickness. For very small values of the wall thickness, the exponent approaches a value of three as expected when the deformation of the panel is completely accommodated by the face sheets and the deformation of the cylindrical ligaments is negligible. As the wall thickness increases, the bending stiffness of the face sheets increases as $t^3$, while the shear stiffness of the cylindrical ligaments increases only linearly. As a result, the core accommodates a larger fraction of the deformation and the value of the exponent decreases with increasing $t$. This effect becomes more apparent if we vary the wall thickness of the ligaments, while keeping the thickness of the face sheets constant (Figure 5e). When the thickness of the ligaments is very small, both ligaments and face sheets deform, and the shear stiffness of the panel increases rapidly with ligament thickness. As the stiffness of the core ligaments increases further, the shear stiffness of the panel reaches a plateau where the shear stiffness is completely determined by the bending stiffness of the face sheets. As a result, the plateau values of the shear stiffness have a nearly perfect cubic dependence on the face sheet thickness.

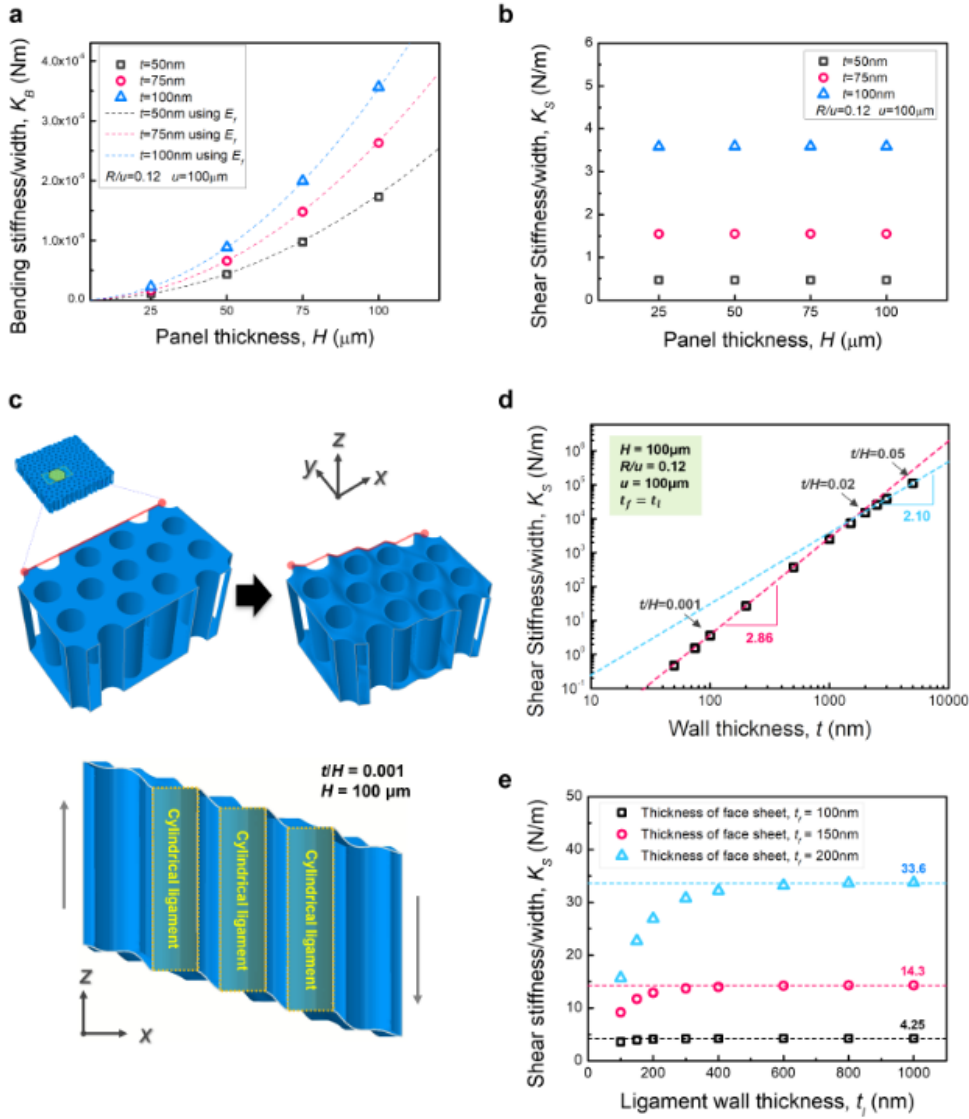

**Figure 5.** For nano-cardboard panels with $u$=100μm and $R/u$=0.12, (a) Bending stiffness for various values of wall thickness and panel thickness; (b) Shear stiffness for various values of wall thickness and panel thickness; (c) Schematic diagram of a panel with a hexagonal ($R/u$=0.12) unit cell subject to simple shear deformation; cross-sectional view of panel along the XZ plane; (d) Plot of shear stiffness as a function of wall thickness. (e) Plot of shear stiffness per unit length as a function of ligament wall thickness for several values of the face sheet thickness.

*Dependence on unit cell pattern*

The unit cell defines the shape, spacing, and size of the ligaments connecting the two face sheets in a nano-cardboard panel. As such, it determines the stiffness of the core and has a profound impact on the overall mechanical behavior of the panel. Figures 6a and 6b depict as a function of area density the bending stiffness and the shear stiffness of nano-cardboard panels with the same wall and panel thicknesses, but with two different types of unit cells - one a hexagonal pattern with circle motif, the other the basketweave pattern used by Bargatin *et al* [8]. As the ligament radius ($R$) or the aspect ratio changes ($a/b$), the nano-cardboard panels sweep a range of area densities. As shown in Fig. 6a, the bending stiffness decreases as the size of the motif increases for both types of unit cell. The bending stiffness is determined mainly by the stiffness of the face sheets and the thickness of the nano-cardboard panel. As the area fraction of ligaments increases, the stiffness of the face sheets decreases (*Appendix C*), and so does the bending stiffness of the panel. Note that, for a given area density, the hexagonal unit cell results in a significantly higher bending stiffness than the basketweave unit cell. The trends are different for the shear stiffness, as shown in Figure 6b. As mentioned earlier, shear deformation of a nano-cardboard panel is accommodated by bending of the face sheets rather than deformation of the core. As the motif size of the pattern increases, the area between the ligaments where the bending deformation concentrates is reduced, resulting in a higher overall shear stiffness. The shear stiffnesses are comparable for both types of unit cell, although the shear stiffness for the basketweave pattern increases more rapidly with area density.

### *3.2. Stiffness of nano-cardboard cantilever beams*

When a transverse load is applied to the end of a typical cantilever plate, the deflection of the plate is predominantly the result of bending deformation. The contribution of shear deformation is usually small except in the case of very short plates. This is not the so for nano-cardboard panels. Because the shear stiffness of a nano-cardboard panel is determined by the bending stiffness of its face sheets, its shear stiffness is relatively speaking much *smaller* than that of regular sandwich panels. As a result, its deflection behavior is different from the behavior of a regular sandwich panel in the sense that shear has a much larger contribution to the overall deflection of the nano-cardboard panel. Consequently, the deflection of nano-cardboard panels cannot be described by Kirchoff-Love plate theory, and Uflyand-Mindlin plate theory [20, 21] should be used instead. According to Uflyand-Mindlin theory, the maximum deflection of a cantilever panel subject to a constant line force $F$ along its edge is the sum of the bending and shear contributions:

$$\delta_{total} = \delta_{bending} + \delta_{shear} = \frac{FL^3}{3K_B} + \frac{FL}{K_S}, \qquad \text{Eq. (2)}$$

where $L$ the cantilevered length of the panel. Note that for this special case the same result can also be obtained from Timoshenko beam theory [19]. Combining both stiffnesses into a single apparent bending stiffness per unit width $D_{app}$, we write

$$\delta_{total} = \frac{FL^3}{3D_{app}} \quad \text{with} \quad D_{app} = \frac{K_B}{1+\frac{3K_B}{L^2 K_S}}. \qquad \text{Eq. (3)}$$

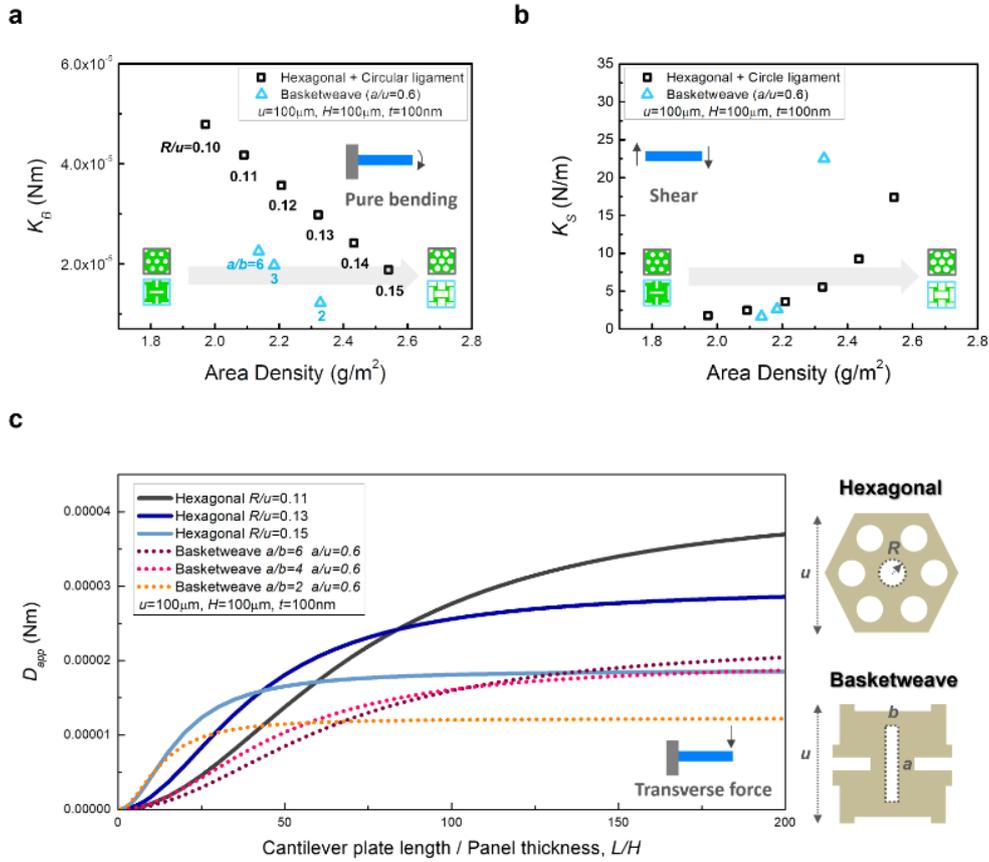

**Figure 6.** (a) Graph of bending stiffness per unit length as a function of area density for nano-cardboard panels with the same shell thickness, but different unit cells; (b) Graph of shear stiffness per unit length as a function of area density for nano-cardboard panels with the same shell thickness, but different unit cells; (c) Plot of apparent bending stiffness calculated using values of the bending stiffness and shear stiffness from FEA simulation for various hexagonal and basketweave patterns ($t = 100$ nm, $H = 100$ μm).

Figure 6c shows a graph of the apparent bending stiffness normalized by $K_B$ as a function of the cantilever length normalized by the panel thickness for various unit cells. The apparent bending stiffness in this figure was calculated using the bending stiffness and the shear stiffness values shown in Figures 6a and 6b. Note that the absolute values of the apparent bending stiffness for panels with a hexagonal unit cell are much larger than for panels with a basketweave pattern and comparable area density. The apparent bending stiffness changes with the length of the cantilever panels because the relative contributions of shear and bending deformation change with panel length (Figure 6c). Because the shear stiffness is so small, the transition from shear to bending occurs at a length to panel thickness ratio that is much larger than observed for a typical macroscopic panel - for a nano-cardboard panel with a length that is 50 times larger than the panel thickness, deformation by shear remains important. This would be highly unusual for a regular sandwich panel.

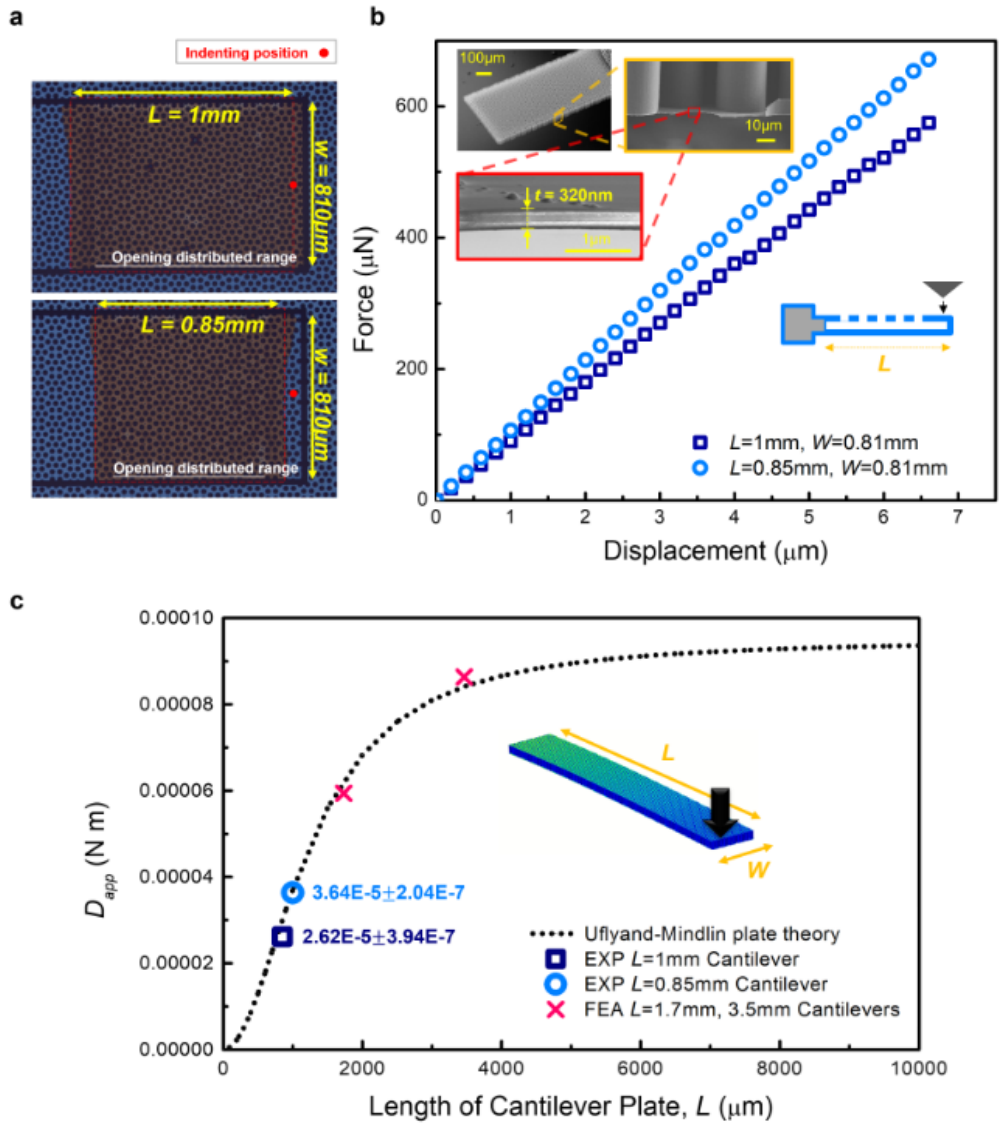

**Figure 7.** (a) Optical image of nano-cardboard cantilever plates of $L$=1000μm and 850μm with $w$=810μm; (b) Force-displacement curves as results of nanoindentation tests for $L$=1000μm and $L$=850μm 2D nanolattice cantilever plates; (c) Plot of apparent bending stiffness obtained from experiments and FEA simulation of the hexagonal pattern cantilever plates ($R/u$=0.13, $u$=100μm, $H$=100μm, $t$=320nm).

### 3.3. Experimental verification

To verify the finite element results experimentally, we fabricated two cantilever panels with lengths 850 μm and 1000 μm out of nano-cardboard, both with the same hexagonal unit cell ($u$ = 100 μm, $R$ = 13 μm), panel thickness (100 μm), and wall thickness (320 nm) (Figure 7a). The area within the red dotted line consisted of nano-cardboard, where the silicon was etched away selectively by making small openings in the aluminum oxide; the area outside the dotted line consisted of solid silicon with a coating of ALD aluminum oxide. The force-displacement curves obtained using the nanoindenter are shown in Figure 7b. The force-displacement curves are linear over the entire range of displacements and a linear least-squares fit of the data allows experimental determination of the values of the apparent bending stiffness $D_{app}$ of the plates using Eq. (3). These experimental values obtained from multiple deflection experiments are shown as blue square and circle in Figure 7c, which graphs the apparent bending stiffness as a function of cantilever plate length. Also marked in this figure are the values of the apparent bending stiffness obtained from FEM simulations of nano-cardboard cantilever plates of slightly different lengths. The black curve is the apparent bending stiffness calculated from Eq. (3) using the values of the bending stiffness $K_B$ and the shear stiffness $K_S$ obtained from finite element simulations of the unit cell models (Figs. 5a and 5b). It is evident from Figure 7c that there is very good agreement between the experimental values of the apparent bending stiffness, the values obtained from the full-fledged cantilever plate simulations, and the curve obtained from Eq. (3). This agreement suggests that Uflyand-Mindlin plate theory provides a good description of the mechanical behavior of cantilever nano-cardboard plates, provided the bending and shear stiffness values are known. The agreement also indicates that the geometry and dimensions, as well as the material properties, of the experimental cantilevers are very close to those of the FEM, i.e., the current fabrication process provides excellent control over the dimensions, form, and properties of the nano-carboard structures.

### 3.4. Post-buckling behavior of 2D nanolattice cantilever beams

The load capacity of sandwich panels is generally limited by various failure mechanisms including face sheet yielding, core yielding, face sheet wrinkling, and buckling [22]. Nano-cardboard panels have an extremely small ratio of face sheet thickness to panel thickness - a typical ratio can be as small as 1:1000. Furthermore, they are fabricated using materials that tend to have very large yield or fracture strengths [23, 24], either because of their very fine microstructures in the case of metals or because they are dense ceramics without macroscopic flaws. Consequently, microfabricated panels buckle easily under load, but the buckling deformation is often entirely elastic and fully reversible [7, 8]. Thus, the non-linear post-buckling behavior of these panels may potentially be regarded as part of their normal useful behavior and quantification of the buckling behavior may enable creative use of these panels in applications.

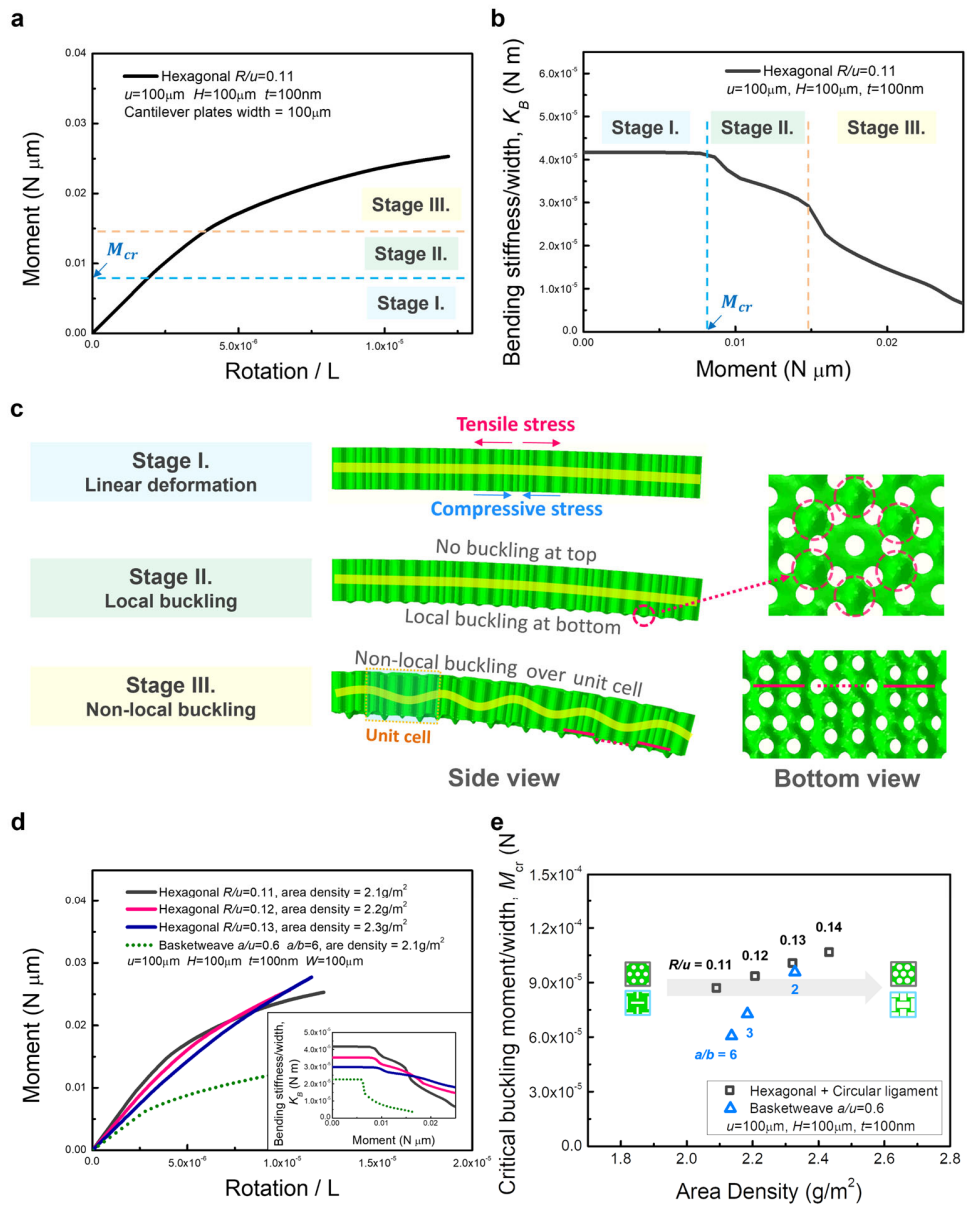

**Figure 8.** (a) Moment-rotation curve obtained from an FEM analysis of pure bending of a typical nano-cardboard panel; (b) Typical bending stiffness-moment curve from pure-bending of nano-cardboard panel; (c) FEA snapshots of buckling phase of nano-cardboard with hexagonal pattern ($R/u$=0.11); (d) Buckling curves of nano-cardboard panels of different patterns with instantaneous stiffness inset; (e) Graph of the critical buckling moment as a function of area density for nano-cardboard panels with the same shell thickness, but different unit cells.

Figure 8a shows a typical moment-rotation curve of a nano-cardboard cantilever with a hexagonal pattern loaded in pure bending obtained by finite element analysis; Figure 8b shows the corresponding instantaneous bending stiffness as a function of rotation. Three stages are clearly visible: In stage I, before buckling, the deformation behavior is linear. In this stage, compressive stresses develop in the bottom face sheet, and tensile stresses in the top face sheet (Figure 8c). In stage II, the compressive stress in the bottom sheet leads to local buckling of the face sheet as shown in Figure 8c. Buckling occurs in the largest areas (red dotted circle in Figure 8c, Stage II) unconstrained by the cylindrical ligaments. While local buckling is accompanied by a reduction of the bending stiffness of the panel, the panel retains significant resistance to bending after buckling (Figure 8b). As the bending moment is increased further, the panel buckles over a length scale that spans several unit cells (Stage III), and the bending stiffness decreases rapidly (Figure 8b and c).

Figure 8d shows the buckling curves for nano-cardboard cantilever panels with various hexagonal unit cells loaded in pure bending; Figure 8e summarizes the effects of ligament radius *r* and area density on the critical buckling moment. The values of the critical buckling moment were obtained from buckling simulations performed using ABAQUS. The first eigenvalue was extracted from the FEM displacement using a linear perturbation analysis and the critical buckling moment was then calculated using the bending stiffness of the structure. As the radius of the cylindrical ligaments increases, the size of the unconstrained area between the ligaments decreases and the critical bending moment for local buckling increases. While the initial bending stiffness of the panel decreases with increasing ligament radius (Figure 8(d)), the reduction in bending stiffness after buckling decreases, i.e., buckling has only a small effect on the load carrying ability of the panel. The figures also show the typical buckling behavior of a panel with a basketweave pattern. In general, the buckling behavior of the basketweave pattern is inferior to that of a hexagonal pattern with similar area density (Figures 8d and 8e). There are two causes for the inferior buckling behavior of the basketweave pattern: 1) The ligaments of the basketweave pattern consist of flat rectangular plates, which buckle at much lower loads than the cylindrical ligaments in the hexagonal pattern; 2) Unlike panels with a hexagonal pattern, basket weave panels loaded in pure bending develop compressive stress in both top and bottom face sheets, causing buckling in the top face sheet in addition to buckling in the bottom face sheet.

Figure 9a shows the instantaneous bending stiffness of nano-cardboard panels with the same hexagonal unit cell but different panel thicknesses, 25 μm and 50 μm, as a function of applied bending moment per unit panel thickness. While the bending stiffness increases with panel thickness, we note that the critical buckling moments per unit panel thickness are approximately the same because the stress distributions in the face sheets scale directly with the applied bending moment per unit panel thickness $M/H$. Figure 9b depicts the region of the face sheets where local buckling occurs in the FEM models when the panel is subjected to pure bending. Buckling occurs in the exact same locations in the FEM models as observed experimentally (inset in Figure 9b) and the buckled shapes are qualitatively similar. Figure 9c shows for various unit cells the normal stress distribution in the *X*-direction along the white dotted line in Figure 9b immediately prior to buckling. It is evident from the figure that nano-cardboard panels with the same unit cell ($R/u$=0.12, 0.13), but different panel thickness, have very nearly the same stress distribution and hence the same $M/H$ immediately prior to buckling, in agreement with Figure 9a. The stress distributions go through a compressive maximum in the region where buckling is observed. There are regions near the edges of the distributions that have higher compressive stresses, but these regions are much more constrained by the ligaments and no buckling is observed. Figure 9d shows the critical value of $M/H$ as a function of $R/u$. As the value of $R/u$ increases, the region where buckling is observed is increasingly constrained by the ligaments and the critical value of *M/H* increases. Figures 9e and 9f show the effects of wall thickness *t* and unit cell size *u* on local buckling of the face sheet. The critical value of *M/H* scales with $t^3/u^2$, similar to the critical buckling condition for a simply supported plate. The critical moment for local buckling of the face sheet can then be estimated from

$$M_{cr} = f(R/u) H \frac{t^3}{u^2} \frac{E}{1-v^2}, \qquad \text{Eq. (4)}$$

where *E* and *v* are Young's modulus and Poisson's ratio of the coating material, and the function *f* (*R/u*) can be obtained from Figure 9d. Quantification of the post-buckling behavior of the panel (Figure 8d) of course requires detailed large-deformation FEM calculations.

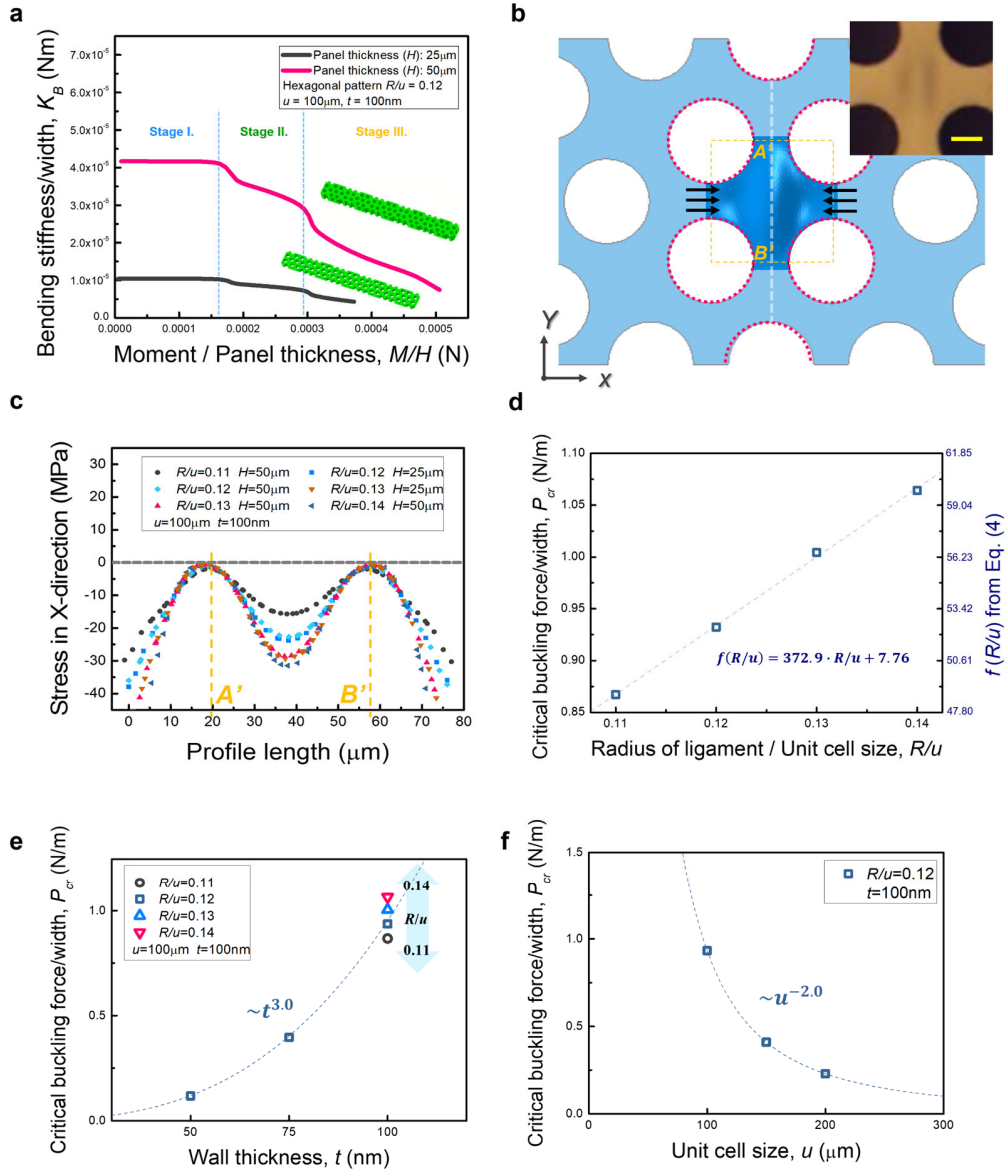

**Figure 9.** (a) Bending stiffness versus applied moment divided by panel thickness for nano-cardboards with various panel thicknesses; (b) Buckled shape of the largest unconstrained section of the bottom face sheet obtained from FEA analysis along with an optical image of a buckled nano-cardboard sample; (c) Stress distribution along the white dotted line of Figure 9b on bottom face sheet before local buckling; (d) Critical buckling force dependance on $R/u$; (e) Effect of shell thickness and pattern on critical buckling force; (f) Effect of unit cell size on critical buckling force. The scale bar in (b) corresponds to a length of 10 μm.

## 4. Some design considerations for nano-cardboard panels

With applications involving photophoretic levitation in mind, we focus this section on developing design principles for nano-cardboard panels that are both ultralight and ultra-stiff. The area density of a nano-cardboard panel with a hexagonal pattern is given by

$$\rho_{area} = \frac{2\sqrt{3} \cdot u^2 t_f + 28 \cdot \pi R(Ht_l - Rt_f) - 14\pi H t_l^2}{\sqrt{3} \cdot u^2} \rho_c, \qquad \text{Eq. (5)}$$

where $\rho_c$ is density of the wall/coating material, $t_l$ is the wall thickness of the ligaments and $t_f$ is the thickness of the face sheets. As shown in Table 1 for the special case $t = t_f = t_l$, low weight and good mechanical performance are inherently conflicting requirements. Any change to $H$, $t$ or $u$ that improves the mechanical behavior also results in an increase in area density (Eq. (5)). It is possible, however, to improve the mechanical performance and keep the area density constant by increasing $t$ and $u$ simultaneously. For instance, a nano-cardboard panel with $u$ = 100 μm, $H$ = 100 μm, $t$ = 100 nm, and $R/u$ = 0.12, has an area density of 2.20 g/m². If $t$ is doubled and $u$ multiplied by a factor of 2.53, the area density of the panel remains the same, while $K_B$, $M_{cr}$, and $K_S$ improve slightly. Increasing the value of $u$, however, also increases the radius $R$ of the

Table 1. Relationship between the mechanical behavior and area density of hexagonal pattern nano-cardboard panels, and structural parameters for the case that $t_f = t_l$.

| | Panel thickness (H) | Wall thickness (t) | Unit cell size (u) | Ligament radius / Unit cell size (R/u) |
|---|---|---|---|---|
| **Bending stiffness, $K_B$** | ~$H^2$ | ~$t$ | No effect * | ↓ |
| **Critical buckling moment, $M_{cr}$** | ~$H$ | ~$t^3$ | ~$1/u^2$ | ↑ |
| **Shear stiffness, $K_S$** | No effect | ~$t^3$ | ~$1/u^2$ * | ↑ |
| **Area density, $\rho_{area}$ (Eq. 5)** | ↑ | ~$t$ | ↓ | ↑ ($H > R$) <br> ↓ ($H < R$) |

*Effect of unit cell size ($u$) is specified in *Appendix D*

ligaments, which may have a deleterious effect on the photophoretic force if $R$ becomes larger than the mean free path of the surrounding gas [4].

More headway can be made by allowing the thickness of the face sheets and the ligaments to vary independently since the face sheets and ligaments affect the overall mechanical behavior of the nano-cardboard panel very differently. Different thicknesses for the face sheets and the ligaments are readily implemented in the current fabrication process.

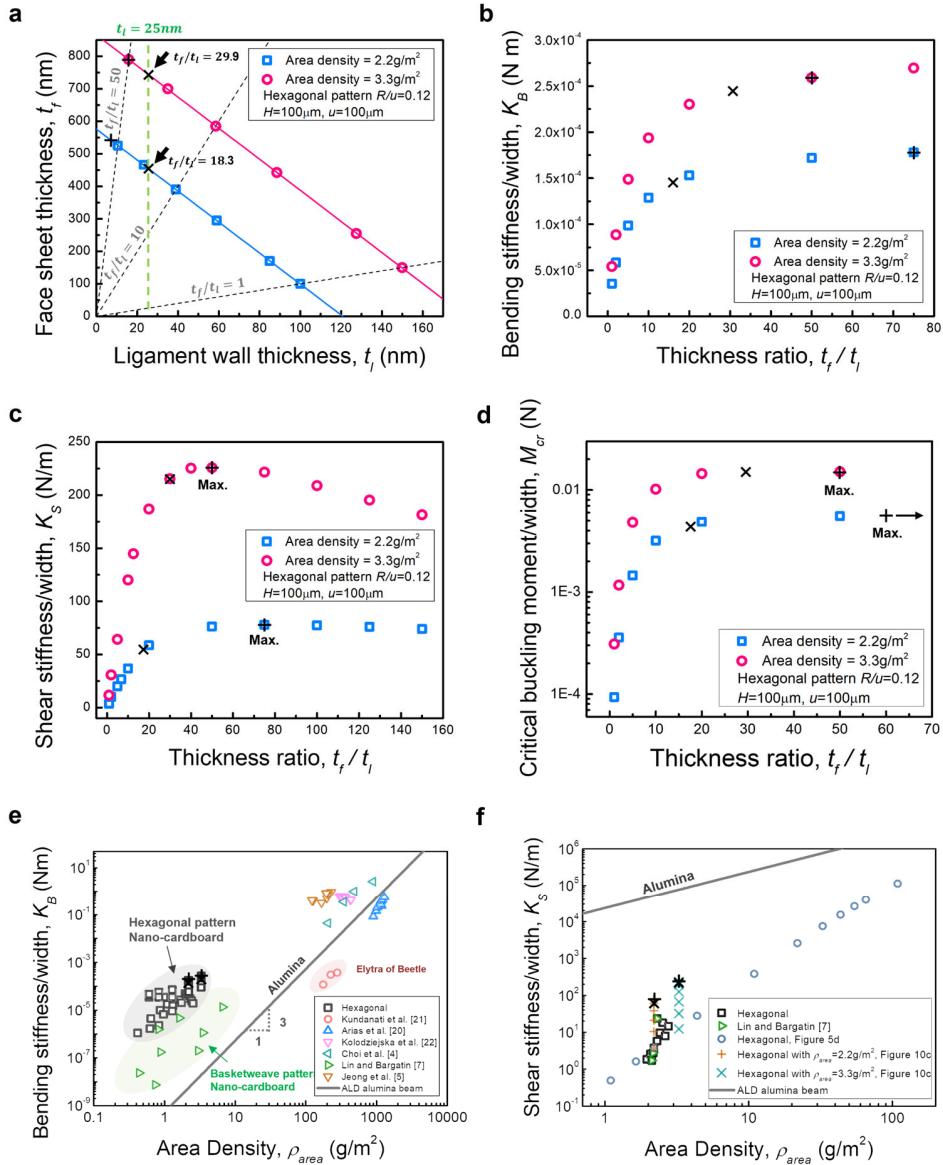

**Figure 10**. (a) Combinations of face sheet thickness and ligament thickness for fixed area densities; (b) Effect of $t_f/t_l$ on the bending stiffness at constant area density; (c) Effect $t_f/t_l$ on shear stiffness at constant area density; (d) Effect of $t_f/t_l$ on the critical buckling moment at constant area density; (e) Bending stiffness of various sandwich structures versus area density; (f) Shear stiffness of various sandwich structures versus area density. In (a)-(d), the hexagonal nano-cardboard panels have dimensions of $H$=100μm, $u$=100μm and $R/u$=0.12.

As shown in Figure 10a, there are many combinations of $t_f$ and $t_l$ that result in the same area density. Figures 10b-d show the values of $K_B$, $K_S$ and the critical buckling moment as a function of the thickness ratio, $t_f/t_l$, for a fixed area density. The figures show that both the bending stiffness and the critical buckling moment initially rise rapidly and then approach a plateau (Figures 10b, d) as the thickness ratio increases for a fixed area density. The reason for this behavior is that both the bending stiffness and the critical buckling moment are determined by the face sheet thickness, and the face sheet thickness approaches a constant value determined by the area density as the thickness ratio grows without bound (Figure 10a). By contrast, the shear stiffness is determined by the thickness of the face sheet when the ligament wall thickness is large and the deformation is localized in the face sheets, but it decreases rapidly with ligament wall thickness when the wall of the ligaments is thin and the compliance of the core starts to dominate the deformation (Figure 5e). Consequently, for a given area density, there exists an optimum value of the thickness ratio that maximizes the shear stiffness for that area density (Figure 10c). Below the optimum value, the shear deformation is mainly accommodated by the bending of the face sheets; above that ratio, it is accommodated by the deformation of the core. Because the apparent bending stiffness of a nano-cardboard panel depends on both $K_B$ and $K_S$, the optimum thickness ratio also results in approximately the maximum apparent bending stiffness. The cases corresponding to the optimum thickness ratios are marked in Figure 10a-d by the + symbols.

The results in Figures 10b-d suggest that the wall thickness of the ligaments should be significantly smaller than the thickness of the face sheets for optimum mechanical performance. There are, however, practical considerations that may put a lower bound on the thickness of the ligament walls from a materials and fabrication point of view. For instance, if the panel is fabricated out of alumina using an ALD process, our experience and previous work [8] suggest a lower bound of approximately 25 nm. This lower bound may be different for other materials and fabrication processes. The corresponding maximum practical thickness ratios are marked in Figure 10a-d by the × symbols. These ratios obviously result in slightly lower than optimum stiffness values.

Figures 10e and 10f show a compilation of the bending stiffness and shear stiffness for a broad range of nano- and micro-cardboard panels [5, 6, 8, 22, 25, 26]. For the same area density, the bending stiffness of nano-cardboard panels is superior to that of a bulk aluminum oxide cantilever beam by many orders of magnitude (grey line in Figure 10e) and the bending stiffness of panels with a hexagonal pattern is also significantly better than that of panels with a basketweave pattern. Nano-cardboard panels with the maximum practical thickness ratio and the optimum thickness ratio are marked in the figure with black × and + symbols, respectively, and clearly have superior bending stiffness. Unlike the bending stiffness, however, the shear stiffness of nano-cardboard panels is smaller than that of alumina beams with the same area density (Figure 10f). As the area density increases, the gap between cardboard panels and bulk alumina decreases, but there is a difference of approximately three orders of magnitude in the relevant nano-cardboards region (<10 g/m$^2$). This small shear stiffness affects the bending behavior of nano-cardboard panels, especially for narrow panels as shown in Figure 6c.

This study focuses on the mechanical behavior of nano-cardboard for use in photophoretic levitation devices. For levitation, the temperature difference between top and bottom face sheets needs to be maximized to increase the photophoretic force. To achieve this, Schafer et al. suggested a modified nano-cardboard structure by removing some ligaments to decrease thermal conduction through the ligaments [4]. As the temperature difference between the face sheets of this structure is further maximized, the lofting force is greatly improved by thermal transpiration through the openings in the top and bottom face sheets. The lack of ligaments, however, deteriorates the buckling resistance and stiffness of the structure. We believe that the best photophoretic levitation device consists of two distinct parts, a photophoretically active structure with fewer ligaments and the nano-cardboard covered in this study to provide mechanical stability to the device. It is necessary to study practical photophoretic levitation devices under conditions that optimize both the photophoretic force and the mechanical behavior. That will be the follow-up work of this study.

## 5. Conclusions

We have developed a class of ultralight and ultra-stiff nano-cardboard panels using a scalable fabrication process. The cores of the panels consist of hexagonal patterns of tubular ligaments and results in superior mechanical properties compared to previously reported basketweave patterns. The fabrication process uses a silicon wafer as a sacrificial template and ALD aluminum oxide as structural material. Small openings in the ALD coating distributed across the template make it possible to etch the sacrificial template in selected areas and to scale the size of the panel without impact on etch time. Nano-cardboard panels as large as 3x3 cm$^2$ have been fabricated in just a few hours using this process. We have evaluated the mechanical behavior of the panels using finite elements and find very good agreement of the model results with experiment measurements of the bending stiffness of cantilevered panels. Both experimental and computational results indicate that the panels are well described by Uflyand-Mindlin plate theory, which can account for the relatively low shear stiffness of the panels. We show that for a given area density, there exists a ratio of face sheet thickness to ligament wall thickness that maximizes the shear stiffness of the panel. Below this optimum value, the shear deformation is accommodated by bending of the face sheets; above that ratio, it is accommodated by deformation of the core. Panels with the optimum thickness ratio significantly outperform other panels with the same area density. We believe that these panels are light and strong enough to serve as components in photophoretic levitation devices.

## 6. Acknowledgements

This research was supported by the Star Friedman Challenge for Promising Scientific Research at Harvard University and by the Harvard University MRSEC, which is funded by the National Science Foundation under Grant DMR-2011754. It was performed in part at the Harvard University Center for Nanoscale Systems (CNS), which is supported by the National Science Foundation under NSF ECCS award No. 1541959.

*Appendix A: Isotropic hexagonal pattern*

For the nano-cardboard panel to be elastically isotropic, we suggest the use of cylindrical ligaments organized in a hexagonal pattern based on a honeycomb. Unlike the basketweave pattern, this pattern results in isotropy because of its six-fold rotational symmetry. As an example, we apply pure bending to two hexagonal nano-cardboard plates ($R/u$=0.12, $u$=100μm, $H$=100μm, $t$=100nm). Two cantilever plates were modeled by extending a square hexagonal pattern unit cell in the direction of the X-axis and Y-axis (Figure A1a), respectively. Periodic boundary conditions were applied in the direction perpendicular to the extension axis. Figure A1b shows the moment-rotation curves for both plates, and the overlap between the curves is evident.

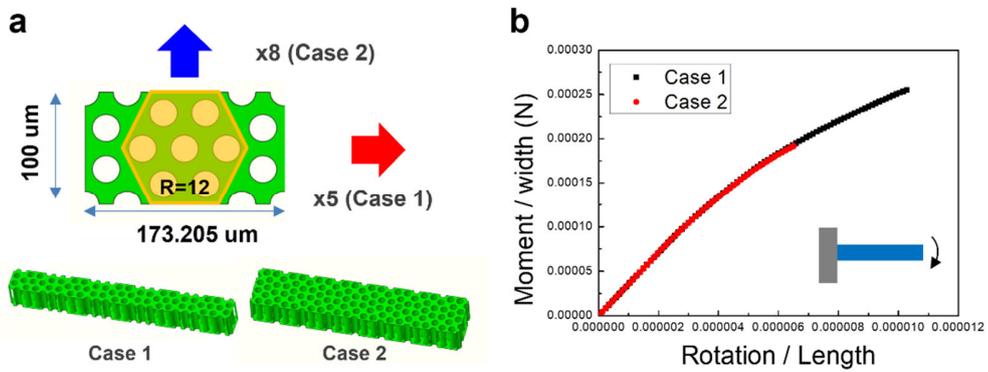

**Figure A1.** (a) Nano-cardboard plates based on hexagonal pattern ($R/u$=0.12, $u$=100μm, $H$=100μm, $t$=100nm); (b) Moment-rotation curve from pure-bending of nano-cardboard plates.

*Appendix B: Effect of micro-openings on bending behaviors*

      Micro-openings in the face sheets dramatically reduce the time for etching silicon using $XeF_2$ gas during the nano-cardboard fabrication. The radius of the micro-openings is 1-2μm, which is small compared to the dimensions of nano-cardboard panel. Using finite element analysis, we checked the effect of those micro-openings on the pure-bending deformation of a nano-cardboard panel. We modeled nano-cardboards without micro-openings and with micro-openings as shown in Figures B2a and b. For the nano-cardboard with micro-openings, we applied pure bending in the downward and upward directions, respectively. As can be seen in Figures B2c and d, the behaviors do not change significantly when micro-openings are present, although the critical buckling moment may be slightly reduced.

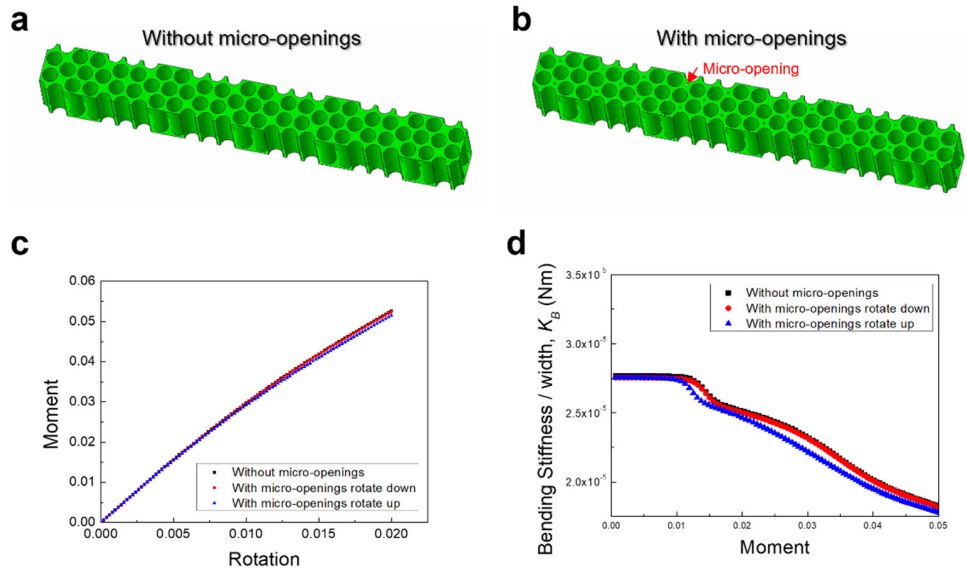

**Figure B2.** Nano-cardboards ($R/u$=0.14, $u$=100μm, $H$=100μm, $t$=100nm) (a) without micro-openings and (b) with micro-openings; (c) Moment-rotation curves from pure-bending of nano-cardboard plates; (d) Bending stiffness-moment curve from pure-bending of nano-cardboard plates.

*Appendix C: Linear sandwich theory for bending stiffness*

Sandwich structures are generally composed of two face sheets and a core connecting them as shown in Figure C3a. Using the parallel-axis theorem, the equivalent flexural rigidity of the sandwich structure ($EI_{eq}$) can be derived as following:

$$EI_{eq} = \sum E_i I_i = E_f(I_1 + I_2) + E_c I_c$$

$$= E_f \left( \left[\frac{wt^3}{12} + wt\left(\frac{H}{2}\right)^2\right] + \left[\frac{wt^3}{12} + wt\left(\frac{H}{2}\right)^2\right] \right) + E_c \frac{w(H-t)^3}{12}$$

$$= E_f \left(\frac{wt^3}{6} + \frac{wtH^2}{2}\right) + E_c \frac{w(H-t)^3}{12}.$$

For nano-cardboards panels, $E_c/E_f$ is approaches zero and the equation above can be expressed as

$$EI_{eq} = E_f \frac{wtH^2}{2}.$$

In nano-cardboard panels, $E_f$ can be obtained from the in-plane tensile deformation of the whole structure including the core (Figure C3b). Since the in-plane deformation of face sheets is somewhat constrained by the core ligaments, the value of $E_f$ is slightly different from the effective elastic modulus of just the patterned face sheet. Figure C3c shows $E_f$ values for various hexagonal patterns, where we used a Young's modulus of 170 GPa and a Poisson's ratio of 0.21 in the finite element analysis. This result also shows that $E_f$ decreases with increasing area fraction of the ligaments.

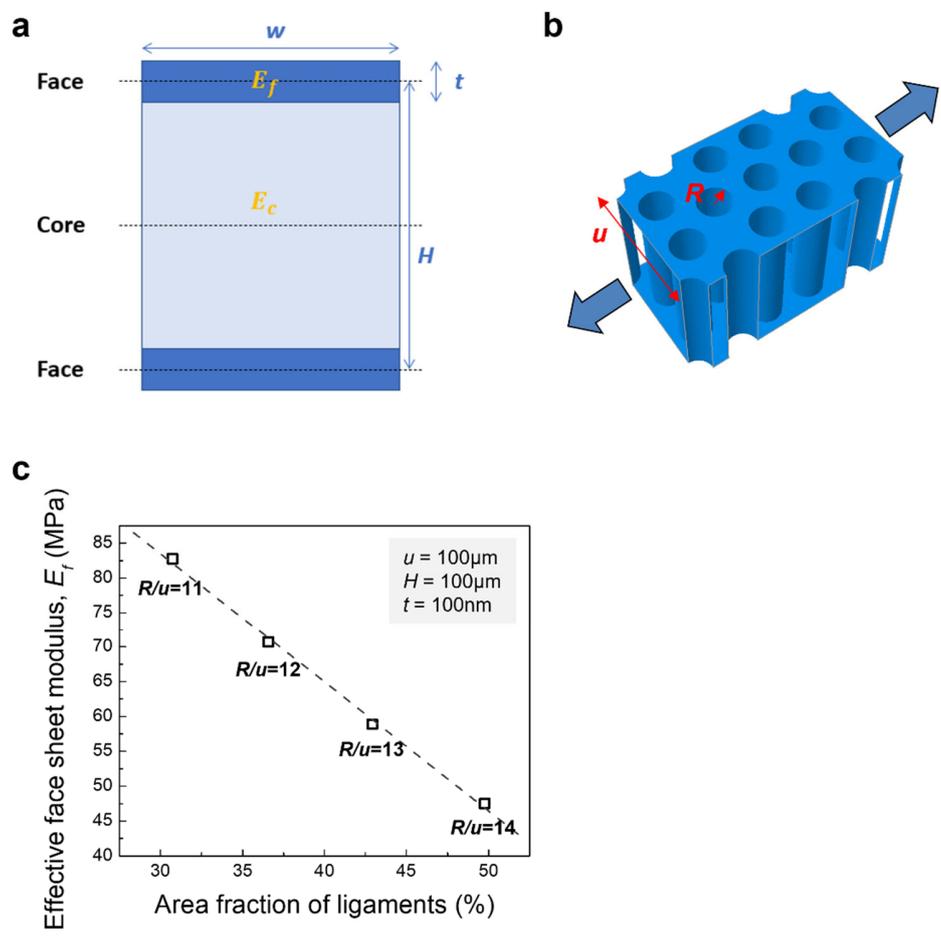

**Figure C3.** (a) Cross-section of typical sandwich structure; (b) In-plane tensile deformation of nano-cardboard structure; (c) Effect of area fraction of the ligaments on the effective face sheet modulus of nano-cardboard panels with hexagonal patterns

*Appendix D: Effect of unit cell size on stiffness of nano-cardboard*

Figure D4 shows the change of bending stiffness and shear stiffness depending on the unit cell size for hexagonal nano-cardboard panels. Using ABAQUS, we prepared nano-cardboard panels with $H$=100μm, $t$=100nm, and $R/u$=0.12. We simulated the bending and shear behaviors for various unit cell sizes ($u$=100, 200, 400, 800μm). As shown in Figures D4(a), the bending stiffness of the nano-cardboard panels is independent of the unit cell size of the panel. The results in Figure D4(b) suggest that the shear stiffness of nano-cardboard panels is approximately proportional to $u^{-2}$.

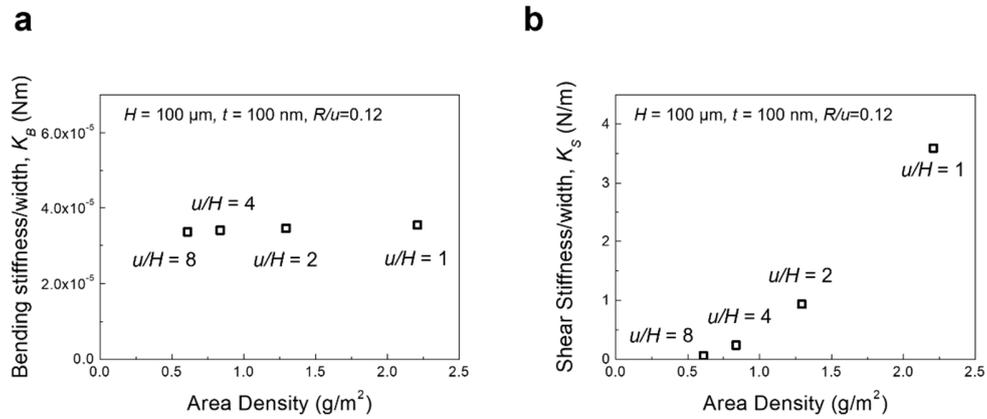

**Figure D4.** For nano-cardboard panels with $H$=100μm, $t$=100nm and $R/u$=0.12, (a) Bending stiffness for various values of unit cell size; (b) Shear stiffness for various values of unit cell size.